\documentclass[12pt]{article}

\usepackage{newtxtext,newtxmath}

\usepackage{graphicx}

\usepackage[colorlinks=true, linkcolor=blue, citecolor=blue, urlcolor=blue]{hyperref}

\usepackage[letterpaper,margin=1in]{geometry}

\renewenvironment{abstract}
	{\quotation}
	{\endquotation}

\date{}

\makeatletter
\renewcommand{\fnum@figure}{\textbf{Figure \thefigure}}
\renewcommand{\fnum@table}{\textbf{Table \thetable}}
\makeatother

\usepackage{scicite}

\usepackage{url}

\def\scititle{Emergence of Pascal’s triangle in cascaded polarization optics: an intuitive framework for field transformation}
\title{\bfseries \boldmath \scititle}

\author{
Ata~Ur~Rahman~Khalid$^{1\ast}$,Naeem~ Ullah$^2$, Nannan~Li$^3$, Hui~Li$^3$,\and Muhammad Ali~Babar~Abbasi$^4$, Robert~M~Bowman$^{1\ast}$\and
	\small$^{1}$Centre for Quantum Materials and Technologies, School of Mathematics and Physics,\and \small Queen's University Belfast, University Road, Belfast BT7 1NN, United Kingdom\and
    \small$^{2}$Center for Optical and Electromagnetic Research, College of Optical Science and Engineering, \and \small Zhejiang University, Hangzhou, 310058, China.\and
    \small$^{3}$ College of Integrated Circuits and Optoelectronic Chips, Shenzhen Technology University,\and  \small Shezhen 518118, China\and
    \and \small$^{4}$ School of Electronics, Electrical Engineering, and Computer Science, Queen’s University Belfast, \small \and \small BT9 5BN Belfast, United Kingdom \and
\and \small$^\ast$Corresponding author.~ ataurrahman.khalid$@$qub.ac.uk;~r.m.bowman$@$qub.ac.uk\and
}

\begin{document} 

\maketitle

\begin{abstract} \bfseries \boldmath
{Nature is imbued with mathematics, manifested through its stunning patterns, symmetries, and structures. Here, we unveil that in a multilayered framework of twisted birefringent optical components, a
recursive number pattern of Pascal's triangle is naturally embedded in the structure of the Jones matrix which intuitively provide a generalized solution for pixel-to-pixel field transformation. The resulting standalone solution is universal across the electromagnetic spectrum, unifies N-layered metasurface and conventional bulk waveplates in a single framework, offers comprehensive insights about the bidirectional complex amplitude modulation and wavefront engineering in linear and circular polarization bases, and at the same time substantially reduces the computational cost. In essence, the discovery of number patterns in polarization optics/photonics will have broad impact across quantum optics, theory informed artificial intelligence model trainings, biomedical engineering and imaging, polarization information encryption, and advanced sensing applications.}
\end{abstract}

\noindent

Nature speaks in numbers through its many facets, which can be seen in intricate patterns and symmetries all around us. Such symmetries and patterns are not only aesthetically appealing but also deeply rooted in fundamental laws of nature and physics. Pascal triangle, recursive pattern of numbers, is a mathematical pattern known to be commonly found in the natural world, and well-documented across various fields of science \cite{sendker2024emergence,mitchison1977phyllotaxis,pletser2017fibonacci}. In the field of electromagnetism, does optical structures follow an identifiable number pattern under certain physical conditions?\par

The study of polarization manipulation is a key topic in both fundamental physics and applied optics; however, the underlying behavior of light propagation through the multilayer systems remains largely unexplored. The Jones matrix formalism is a mathematical framework for analyzing the polarization states of light \cite{xiong2023breaking,bao2021toward}; which dictates the transformation of input polarization states into desired output states through the $2\times2$ Jones matrix $J=\left[\begin{array}{cc}J_{11} & J_{12} \\ J_{21} & J_{22}\end{array}\right]$. In bulk optical waveplates (WPs), the uniform surfaces take the fixed complex amplitude values on $J$ channels, limiting their ability to achieve dynamic wavefront modulations. In contrast, artificially engineered sub-wavelength interfaces (metasurfaces) composed of spatially distributed light-scattering meta-elements \cite{yu2011light} enable unprecedented control over the outgoing channels of $J$ by leveraging the local modulation of the refractive index through structural isotropy or an-isotropy and in-plane rotation at individual pixel which has revolutionized polarization manipulations \cite{rubin2021jones,yang2023integrated,dorrah2022tunable,devlin2017arbitrary}. Despite overwhelming control on the outgoing channel, monolayer metasurfaces are fundamentally limited to modulate three out of the four outgoing channels through the propagation phase and/or geometric phase \cite{devlin2017rc} with directionally symmetric optical response (\textcolor{blue}{Supplementary Text}).\par
To surmount the channel capacity limit and enable bidirectional control in metasurfaces, more recently, alternative approach of layers stacking of rotated anisotropic structures is being adopted. Though, the multi-layered framework have demonstrated remarkable capabilities to realize asymmetric response \cite{liu2016high, kim2024bidirectional}, circular dichroism \cite{zhao2022realization}, holographic information encoding \cite{chang2022universal,wei2022rotational,bao2022observation,shi2022nonseparable,deng2022bilayer}, dynamic beam steering and focus tuning \cite{jung2023three,li2024simultaneous,mansouree2020multifunctional,zhang20236g,cai2021dynamically,luo2021varifocal}, and vortex beam generation \cite{mei2023cascaded,zhang2024light,liu2024all,ma2024electrically}; the theoretical foundations behind their operation are still mysterious. Till to date, only few efforts have targeted bi-layer \cite{deng2023full,chang2022universal} and few-layer metasurface configurations \cite{yuan2020independent,yuan2024reaching,shi2022nonseparable} to explain the complex amplitude transformation; yet these approaches offer partial solutions and fail to explain the underlying mechanisms of field transformation in multilayered configuration. It is worth noting that, due to relative rotations of anisotropic meta-elements in parallel configuration, unlike the isotropic structures \cite{zhou2018multilayer,wei2020compact,ogawa2022rotational,zhang2023highly}, the phase modulation of in-plane rotated meta-elements cannot be achieved by simply summing up the phases of meta-elements facing each other. To extract optimal parameters and complex amplitude for the outgoing channels through a cascaded arrangement of metasurfaces, library or pool search is the traditionally employed approach which has also been complemented by advanced optimization techniques including inverse design and deep learning driven by artificial intelligence and machine learning tools \cite{wei2022rotational,huang2023all,georgi2021optical,zhang2025vectorial,sun2024nonclassical,zhang2024dual,li2024simultaneous}. Though, these methods have achieved notable successes; however, they often operate as ``black boxes'', providing empirical solutions without fully elucidating the underlying principles. \par
Studies suggest that intrinsic analogies exist between the bulk and metasurfaces WPs \cite{chen2021highly,liu2017single,deng2022recent} and their cascaded configuration are indispensable in polarization manipulation \cite{vilas2022customized,herrera2015design,georgi2021optical,wei2022rotational}. The complex amplitude calculations for the cascaded configuration of bulk optical WPs are
superficially simple, while they are daunting and resource intensive for the cascaded metasurfaces WPs. This study unifies both and reveals the underlying dynamics and hidden symmetries which regulate bidirectional field transformation. Through rigorous analysis, we discovered that the nature inherently encodes the rotation terms of the Jones matrix elements in a specific arrangement which follow a particular recursive pattern of numbers representing number triangles i.e., Pascal/Al-Samaw'al triangles, marking its first observation in polarization optics. Hence, leveraging from infinite series of numbers, the number patterns assist in formulating the generalized solution for cascading birefringent elements which effectively reduces the computational cost by predicting and identifying the specific parameters
required for desired complex amplitude modulation-and at the same time shed light on multiple fascinating phenomena that were previously misunderstood or not observed.  For instance, unlike the common perception \cite{huang2025chirality}, there exists rotation-induced geometric phase terms in all channels of  circular polarization (CP)  in twisted cascaded configuration, which rely on the relative angular interaction according to the binomial coefficients. The analytical solution indicates that the diagonal and off-diagonal exchange their position differently in linear and CP bases when light propagates in reverse direction. It also highlight that the CP cross channels can be muted by adopting bi-lyaer HWPs metasurface designs. Furthered, for advanced wavefront engineering through metasurfaces, it is more effective to use combination of WPs rather than relying on HWPs, as it enables more efficient utilization of all outgoing channels (\textcolor{blue}{Supplementary Text}). This is not full extent of the work; there is still more to explore from this framework . \par
To validate the analytical solution based on number patterns, we first performed experiments at optical frequency by setting up a generic platform of bulk WPs with stacked and twisted configuration. Next, we extended our work to explore its applicability for metasurfaces across a broader spectrum, particularly at optical and millimeter waves (mmWave). The experimental and numerical results aligned seamlessly with our theoretical analysis, providing robust validation of our analytical solution both for linear polarization and CP. The generalized solution is universal for cascaded isotropic and/or anisotropic metasurfaces and bulk optical WPs, ensuring the design flexibility, control on channel manipulation, reconfigurability, and much more. In this study, our investigation targeted the fundamental aspects and is confined only to the areas discussed above, which is not exhaustive; the scope of the analytical solution can be further expanded for the deeper exploration and discovery in polarization electromagnetism. Furthered, the generalized control of both polarization bases through multiple layers implies a new paradigm for information encryption, high-resolution imaging, data storage, high-capacity display, and quantum technologies.
\section*{Generalized Jones matrix and recursive number patterns in linear basis}
\label{Generalized Jones matrix-Linear}
We firstly delve into the stacked anisotropic cascaded optical system for complex amplitude modulation, and perceive the number patterns in the Jones matrix. By introducing rotation matrix, the Jones matrix for the monolayer rotated birefringent WP in linear basis can be reduced as follows (eq. S2):
\begin{equation}
\begin{array}
{l}J_{Linear}^n=\left[\begin{matrix}t_{p n}+t_{q n} \cos 2 \theta_n\ \ &t_{q n} \sin 2 \theta_n\\t_{qn} \sin 2 \theta_n&t_{p n}-t_{q n} \cos 2 \theta_n\\\end{matrix}\right]
\end{array}
\label{Eq:jones_Outgoing_Matrix}
\end{equation}
Where $t_{pn} =(t_o^n+t_e^n)/2$ and $t_{qn}=(t_o^n-t_e^n)/2$, which plays a key role in energy routing and phase modulation in outgoing channels. The analytical solution of Jones matrix in a rotated cascaded system becomes sequentially complex, and till now there is no mathematical framework that fully accounts for the polarization transformation in the stacked system. In this case, the transformation matrix of passive, lossless, reciprocal, and coupling-free system in forward (F) direction can be obtained by multiplying the respective Jones matrices in the reverse order in which the light encounters the optical elements as illustrated in Fig. \ref{fig:Schematic}A, which can be mathematically expressed as below: 
\begin{equation}
  J^{F-N-layerd}=J^N*J^{N-1}*J^{N-2}* \dots *J^3*J^2*J^1=\prod _{n=1}^N J^{N-n+1} 
\end{equation}
where $J^n$ is the Jones matrix for the $n^{th}$ element encountered by the electromagnetic (EM) wave in the system. Through careful analysis of the Jones matrix $J^{F-N-layerd}_{Linear}$ components in linear basis, it is observed that the mother nature encoded the rotation angles of birefringent structures in a particular pattern named as “Pascal triangle”, which highlights the inherent order of rotation angle terms and symmetry within the system. The coefficients of the angle terms also follow a specific pattern which can be demonstrated through inclusion and exclusion scheme; it further deepens the connection with naturally recursive sequence and provides an exact solution to the polarization problem for the N-layered birefringent optical elements (Supplementary Text \textcolor{blue}{S1}). The compact generalized Jones matrix for the cascaded birefringent WPs system in linear basis can be described through nested sum as below:  
\begin{equation}
\begin{aligned}
J_{Linear}^{F \_N \_Layered} &=  \sum_{\substack{M=0\\ M \leq N}}
\mathbb{A}\begin{bmatrix}  \cos \mathbb{R} & 0 \\ 0 &  \left(-1\right)^M\cos \mathbb{R} \end{bmatrix}+ \sum_{\substack{M=1 \\ M \leq N}}
\mathbb{A}\begin{bmatrix} 0 & \sin \mathbb{R}\\ \left(-1\right)^{M+1}\sin \mathbb{R} & 0 \end{bmatrix} 
\end{aligned}   \label{Eq:general_Linear}
\end{equation}
where ``$\mathbb{A}=\sum_{1 \leq n_{\alpha_1} < \dots < n_{\alpha_M} \leq N}\left( \prod_{n \notin \{n_{\alpha_1}, \dots, n_{\alpha_M}\}} t_{pn} \right) \left( \prod_{j=1}^{M} t_{q{n_{\alpha_j}}} \right)$'' governs the coefficients of the rotation terms ``$\mathbb{R}=\left( 2 \sum_{j=1}^{M} (-1)^{j+1} \theta_{n_{\alpha_j}} \right)$''. The generalized Jones matrix indicates that $N$-layered stacked birefringent optical elements follow a number pattern in angular interaction count ``$M$'' in all Jones matrix elements which can be described by binomial coefficients $C\left(N,M\ \right)=\frac{\left\{N!\right\}}{\left\{M!\left(N-M\right)!\right\}}$. The Al-Samaw’al/trimmed left aligned Pascal triangle ($0 \leq M\leq N$) and trimmed left aligned Pascal/Al-Samaw’al triangle ($1 \leq M\leq N$), Fig. \ref{fig:Schematic}B or tables. \textcolor{blue}{S1} and \textcolor{blue}{S2}, assist in readily picking the exact values of the ``$M$'' for the diagonal and off-diagonal elements of the Jones matrix, respectively (Supplementary Text \textcolor{blue}{S1} and Supplementary Movie \textcolor{blue}{S1}).

In backward (B) direction the EM waves encounter the elements in reverse order as depicted in Fig.\ref{fig:Schematic}a, thus the analytical solution of Jones matrix in ``B'' direction can be obtained by multiplying the respective Jones matrices in the reverse order in which the light encounters the optical elements as below:
\begin{equation}
    J^{B-N-layerd}=J^1*J^2*J^3* \dots *J^{N-2}*J^{N-1}*J^N=\prod _{n=1}^N J^n 
\end{equation}
From the detailed analysis of wave propagation in reverse direction for $N=1-4$ in Supplementary Text \textcolor{blue}{S1}, it is noted that the Jones matrix associated with a single layer optical component exhibits symmetric properties in the ``B'' direction, which keeps the optical response remain identical in both ``F'' and ``B'' cases.  However, in stacked metasurfaces, due to the non-commutative nature of matrix multiplication system symmetry breaks which accounts for asymmetric response. After a careful analysis, we observed that in the ``B'' direction, the Jones matrix follows the same number patterns observed in the ``F'' direction, but the reversal of Jones matrix multiplication operation makes the exchange of off-diagonal elements. The pattern followed in the ``B'' direction confirms the inherent consistency and mathematical elegance which ensures that the optical response remains predictable and analytically tractable even when the direction of propagation is reversed. The Jones matrix in ``B'' direction can be written as: $J^{B-N-layerd}=(J^{F-N-layerd})^T$, where ``T'' is the transpose.
\section*{Generalized Jones matrix and recursive number pattern in CP basis}
The complex amplitude modulation in a rotated cascaded system is further studied in CP basis, where we derived the analytical solution by solving the Jones matrices and observed the particular number pattern (Supplementary Text \textcolor{blue}{S2}). The compact generalized Jones matrix for the cascaded birefringent WPs system in CP basis can be described through nested sum as below:
\begin{equation}
\begin{aligned}
J_{CP}^{F \_N \_Layered} &=  \sum_{\substack{M even \\ M \leq N}}
\mathbb{A}\begin{bmatrix}  e^ {-i \mathbb{R}} & 0 \\ 0 &  e^ {i \mathbb{R} } \end{bmatrix}+ \sum_{\substack{M odd \\ M \leq N}}
\mathbb{A}\begin{bmatrix} 0 & e^{i \mathbb{R}} \\ e^{-i \mathbb{R}} & 0 \end{bmatrix} 
\end{aligned}   \label{Eq:general_CP}
\end{equation}
The rotation arguments of the Jones elements in the general expression of $N$-layered stacked birefringent optical elements in CP basis exhibit a number pattern in a distinct fashion in their diagonal and off-diagonal elements. It is evident from Eq. (\ref{Eq:general_CP}) that exponential rotation arguments with even-$M$ go with diagonal elements and odd-$M$ go with off-diagonal elements of the Jones matrix. This pattern can be readily picked from the Al-Samaw’al/trimmed left aligned Pascal triangle, Fig.\ref{fig:Schematic}b or Supplementary table. \textcolor{blue}{S1}, by selecting the odd-columns (even-M) for the diagonal elements and even-columns (odd-M) for the off-diagonal elements of the Jones matrix which assist in readily picking the exact values of the $M$ (Supplementary Movie \textcolor{blue}{S2}); the detailed analysis is provided in Supplementary Text \textcolor{blue}{S2}). Similar to the linear basis, the rotation terms in CP basis can be set through combinatorics and their coefficients can be adjusted through inclusion and exclusion scheme; which provide a complete intuitive solution. Next, we extended our theoretical calculations in ``B'' direction and after the careful analysis, we observed that the Jones matrix elements also exhibit the same pattern as in the ``F'' direction in CP basis. However, the non-commutative nature of matrix multiplication inherently introduces directionality and breaks the system's symmetry which accounts for the exchange of diagonal element positions in a fixed CP basis; leading to an asymmetric response (Supplementary Text \textcolor{blue}{S2}). The consistency in pattern reinforces the systematic nature of the analytical solution and ensures a coherent framework for addressing the Jones matrix components in a cascaded system.

From Eqs. \ref{Eq:general_Linear} and \ref{Eq:general_CP}, it can  be observed that the rotation dependent terms exist in all four channels. And, in metasurfaces designs, independent quadruplex channel modulation for wavefront engineering is only possible when $N\ge2$. Moreover, the generalized solution is intuitive (an example is presented in Supplementary movie \textcolor{blue}{S3} for $N=4$), elucidates complex relationships, significantly reduces the overall computational complexity, and ensures hassle-free calculation process for $N$-layered stacked birefringent optical elements. As a demonstration, the analytical equations are implemented in Matlab to develop a graphical user interface (GUI) for ``$N$= 2'', both in linear and CP bases, which provide detailed pixel-to-pixel bidirectional complex amplitude information of Jones matrix elements for arbitrary WPs in fraction of second on personal computer  (Data S1 and S2).
\section*{Experimental and numerical simulation results}
\begin{figure} 
	\centering
	\includegraphics[width=0.95\textwidth, height=1.1\textwidth]{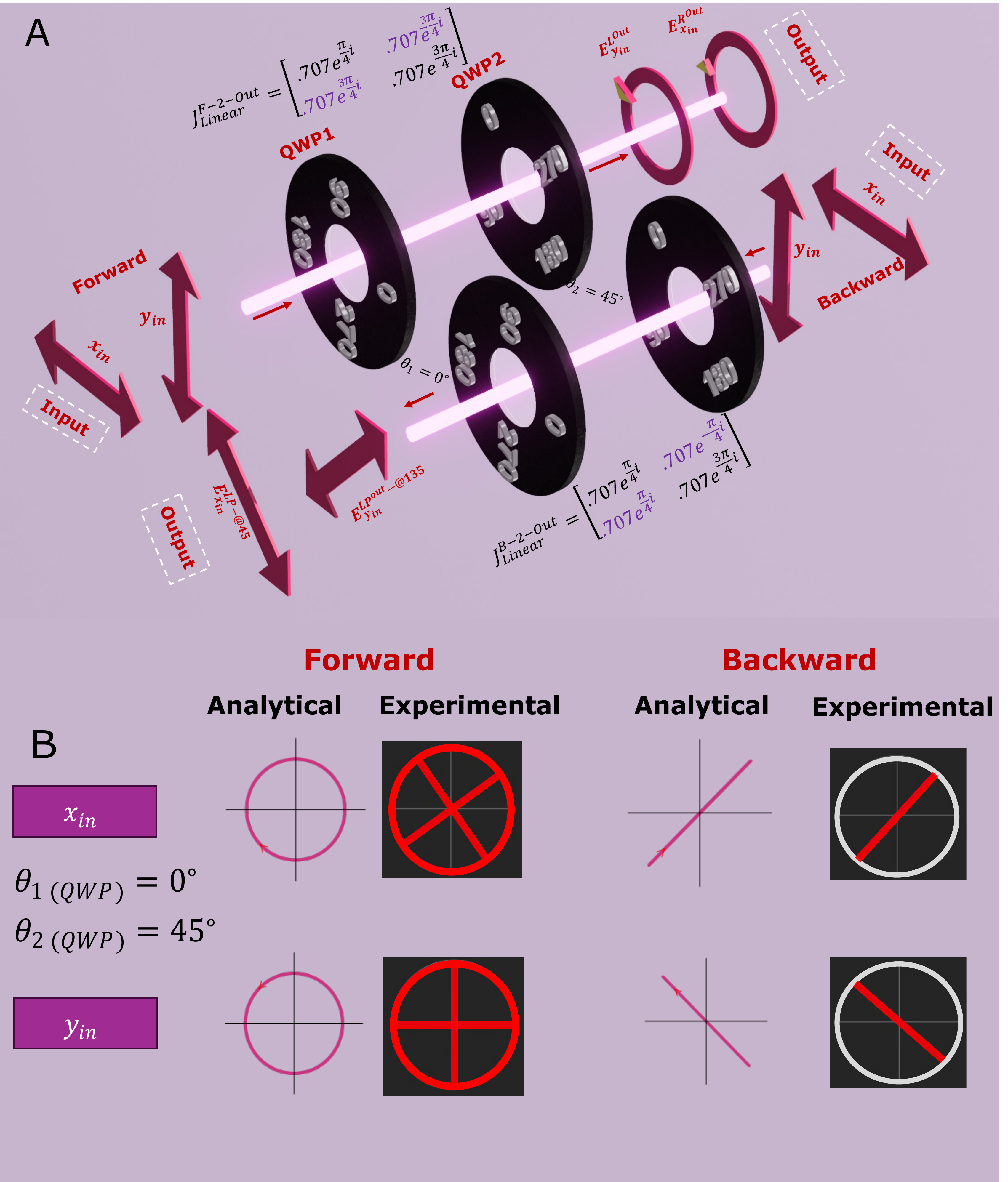} \caption{\textbf{Experimental results validating the analytical solution.} (\textbf{A}) Pictorial illustration of Jones matrix elements swapping and bidirectional polarization conversion in linear basis when light passes through a pair of bulk QWPs orientated at $\theta_1=0^\circ\&\theta_2=45^\circ$. (\textbf{B}) Analytical vs experimental results in ``F'' and ``B'' directions. The experimental results of bidirectional polarization transformation seamlessly match the analytical solution.}
    \label{fig:Main_Exp_Schematic_and_Rlts_2BulkWP}
\end{figure}

To validate the analytical solution based on recursive number patterns, we initially conducted a series of experiments for different cascaded configuration for $N=1-4$ at optical frequency by setting up a generic platform of bulk WPs with stacked and twisted configuration and observe the polarization transformation, which provide foundational insights and confirmed the theoretical (Supplementary Text \textcolor{blue}{S3}); paving the way for extending the concept to more advanced metasurface based designs. In main manuscript, we take account an example of linearly polarized light passes through a set of bulk QWPs as illustrated in Fig. \ref{fig:Main_Exp_Schematic_and_Rlts_2BulkWP}A. By utilizing the GUI (Data S1), we extracted the bidirectional complex amplitude of outgoing Jones matrix elements under the linearly orthogonal polarizations. In ``F'' direction, the complex amplitudes of Jones matrix``$J_{Linear}^{F-2-Out}=.707\left[\begin{array}{cc}e^{i\pi/4} & e^{-i3\pi/4} \\ e^{i3\pi/4} & e^{i3\pi/4}\end{array}\right]$'' dictate that the light converts to CP light whose handedness can be switched by impinging the linearly orthogonal polarized light. The off-diagonal elements of Jones matrix swap their positions by reversal of propagation direction as predicted by analytical model, which aligns precisely with our GUI results as given by ``$J_{Linear}^{B-2-Out}=.707\left[\begin{array}{cc}e^{i\pi/4} & e^{-i\pi/4} \\ e^{i\pi/4} & e^{i3\pi/4}\end{array}\right]$''. The experimental results for the QWPs configuration given in Fig. \ref{fig:Main_Exp_Schematic_and_Rlts_2BulkWP}A validate that bidirectional polarization transformation seamlessly match analytical solution and demonstrate that the Jones matrix element indeed swaps their positions by the reversal of the propagation direction as presented in Fig. \ref{fig:Main_Exp_Schematic_and_Rlts_2BulkWP}B. The polarization transformation results for different cascaded configurations of bulk WPs under linear ``$x_{in}$'' and ``$y_{in}$'' polarization are provided in Supplementary Text \textcolor{blue}{S1}, which also seamlessly match the analytical results; thereby, confirm the accuracy and robustness of analytical solutions.

\begin{figure} 
	\centering
	\includegraphics[width=0.95\textwidth, height=0.9\textwidth]{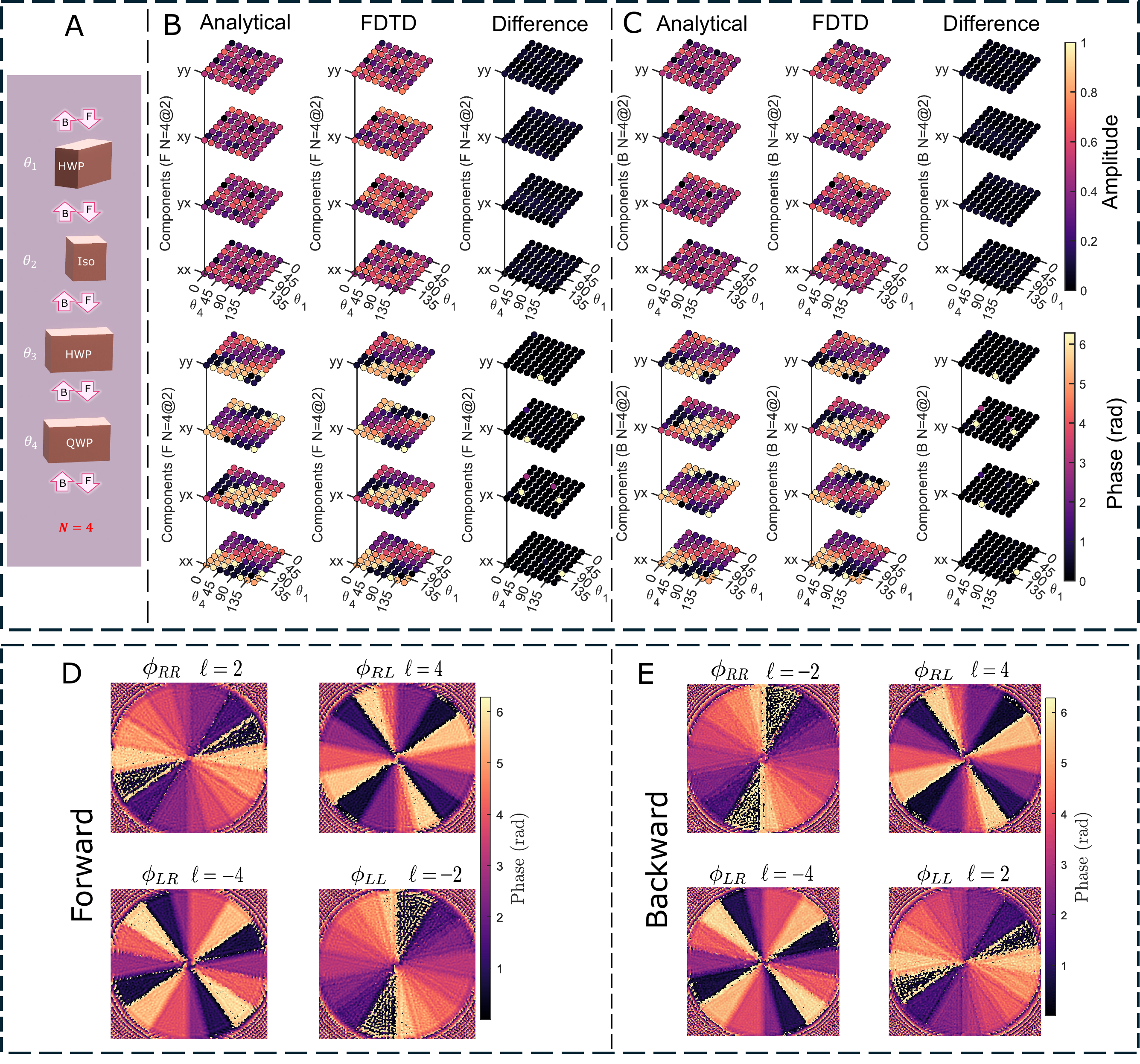} \caption{\textbf{Validation of complex amplitude modulation and bidirectional quadruplex CP channels manipulation based on predicted analytical parameters.} A unitcell configuration of cascaded metasurfaces WPs for N=4. (\textbf{B and C}) Analytical vs numerical (FDTD) results of absolute amplitude, phase and their differences for the linear polarization components when the top and bottom WPs are relatively rotated from $0^\circ$ to $157.5^\circ$ with step size of $22.5^\circ$. The symbol ``@'' indicates that the second layer consists of isotropic meta-element in the unitcell. (\textbf{D}) The nearfield OAM phase profile in `F'' direction and (\textbf{E}) in ``B'' direction. The transmitted phase profiles of the vortex beams with different OAM modes $\ell=2$, $\ell=-4$, $\ell=4$ $\&$ $\ell=-2$ confirms the successful quadruplex OAM generation. The phase profiles of CP outgoing channels in backward direction verify that topological charge values are being exchanged on diagonal elements.}
    \label{fig:Lin_FB_Rlts_N4_Iso_inc_Quadruplex_Phase}
\end{figure}

We extended our investigation to further explore and numerically validate the bidirectional complex amplitude transformation for the dielectric metasurfaces in different cascaded configurations under the linear and CP light for $N=1-4$ (Supplementary Text \textcolor{blue}{S1}). Here, in the main text, we present results for the linearly polarized light impinged on stacked meta-elements arranged in four layers such that the top HWP ($\theta_1$) and bottom QWP ($\theta_4$) rotate relatively, while the middle two layer are fixed to isotropic meta-element ($\theta_2=0^\circ$) and HWP ($\theta_3=135^\circ$), as illustrated in Fig. \ref{fig:Lin_FB_Rlts_N4_Iso_inc_Quadruplex_Phase}A. The analytical and numerical results of bidirectional complex amplitude for the rotation offsets ($\theta_1$) and ($\theta_4$) 
are presented and compared Figs. \ref{fig:Lin_FB_Rlts_N4_Iso_inc_Quadruplex_Phase}, B and C. The comparison reveals that there is no discernible difference between the analytical and numerical results, providing strong and conclusive evidence that analytical model is effective for isotropic and anisotropic geometries and validates the theoretical framework.

To demonstrate the broader applicability of the analytical framework in wavefront engineering, we extended our investigations to validate analytical predictions such as quadruplex independent channel modulation, and asymmetric response in ``B'' direction. The quadruplex independent wavefront manipulation in CP basis has been a research problem for years and only few efforts have been made to realize wavefront engineering on co and cross polarization channels \cite{yuan2020independent,yuan2024reaching,huang2023all,yuan2024full}. These approaches rely on pool search scheme by building extensive library of EM responses in parametric space through numerical simulation. By leveraging our analytical framework, the independent quadruplex CP wavefront manipulation can be conveniently achieved by bi-layer structures consisting of HWP and QWP. Our analytical framework predict and identify the specific parameters required for independent phase manipulation. To do so, we utilize bi-layer cascaded dielectric metasurfaces design and successfully encoded vortex beam carrying orbital angular momentum with distinct topological charge ($\ell$) on individual outgoing channels and observed asymmetric response in ``B'' direction under CP. The phase profiles for different OAM modes, describing the quadruplex channel modulation and asymmetric response, are presented in Fig. \ref{fig:Lin_FB_Rlts_N4_Iso_inc_Quadruplex_Phase}, D and E, while the detailed mechanism and analysis are thoroughly explained in the Supplementary Text \textcolor{blue}{S3}, providing a comprehensive understanding of the design principle. Overall, the generalized solution enables precise multi-layer  complex amplitude control over the outgoing channels, while providing direct physical insights and predictable bidirectional optical response in both polarization bases.

\begin{figure} 
	\centering
	\includegraphics[width=\textwidth,height= 1\linewidth]{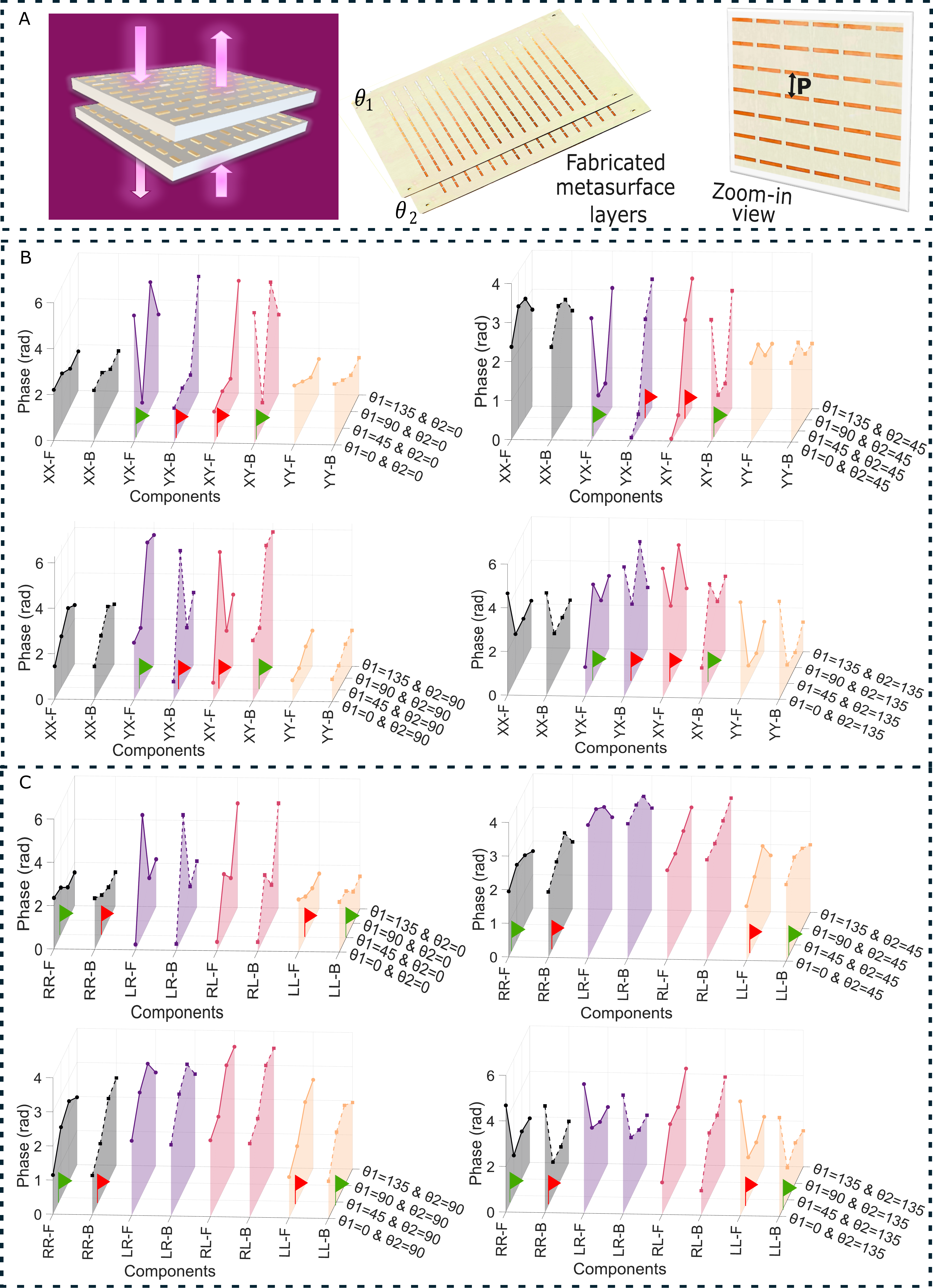} \caption{\textbf{Experimental verification of directionality in linear and fixed CP bases using the bi-layer cascaded metasurfaces designed for mmWave.} (\textbf{A}) The schematic illustration of bidirectional field transformation through non-monolithic bi-layer metasurface design and photographs of fabricated samples with zoom-in view of the proposed design. The ``P'' is the periodicity. (\textbf{B and C}) Bidirectional phase profiles verifying the Jones elements swapping in linear and CP bases. (\textbf{B}) The off-diagonal elements swapped in linear basis and  (\textbf{C}) the diagonal elements swapped in CP basis. The swapped elements are represented by the same color flag marker. The $\theta_1$ and $\theta_2$ are in degrees.}
    \label{fig:mmWave_Exp_Rlts_N2_CP}
\end{figure}

Finally, we conducted experiments to verify the swapping of Jones matrix elements through the phase profiles for cascaded anisotropic metallic meta-elements under the linear and CP polarizations in millimeter wave. Unlike the dielectric structures at shorter wavelength, the $\Delta\phi=\pi$ phase difference between the linearly orthogonal polarization in metallic rod in mmWave is challenging. Also, the unavoidable and undesired factors (reflection/coupling) exist and even amplifies in the stacked metallic configuration in this regime; however, the Jones matrix elements phase profiles swapping condition remains consistent in linear and CP basis when the wave impinges from backside. To validate the analytical prediction of Jones elements swapping, we fabricated two sets of anisotropic rectangular meta-elements array with rotation angles $0^\circ$, $45^\circ$, $90^\circ$ and $135^\circ$ and arranged pairs of meta-elements in a non-monolithic cascaded configuration. The schematic of bi-layer non-monolithic cascaded configuration for bidirectional field transformation, fabricated bi-layer metasurfaces, and zoom-in view of fabricated sample are shown in Fig. \ref{fig:mmWave_Exp_Rlts_N2_CP}A. The bidirectional EM response for all possible bi-layer cascaded configurations was then measured and analyzed. The phase profiles for four different metasurfaces configurations at 13 GHz are shown in Fig. \ref{fig:mmWave_Exp_Rlts_N2_CP}, B and C, where the bottom layers ($\theta_2$) are fixed to $0^\circ$, $45^\circ$, $90^\circ$, and $135^\circ$, respectively, while the top layer ($\theta_1$) are rotated from $0^\circ$ to $135^\circ$ with a step size of $45^\circ$ in each design. 

The experimental results demonstrate the swapping of the off-diagonal Jones matrix in linear basis and the diagonal element swapping in CP basis, validating the inherent directionality of Jones matrices which accounts for the exchange of elements. The corresponding Jones elements swapping is highlighted in the graphs using the same-colored flag markers. In Supplementary Text \textcolor{blue}{S3},  we also included the experimental results of the stacked configuration for three layers consisting of isotropic square patches sandwiched between the rotated rods, which dually verify that the direction-dependent response is consistent for the combination of isotropic and anisotropic geometries. The device fabrication and measurement techniques are discussed in methods. \cite{methods}.
\section*{Conclusion and outlook}
The emergence of Pascal's triangle in polarization optics/photonics provides the meaningful insights, reveals the hidden symmetry, and regulates the mechanism of field transformation in cascaded birefringent optical elements. More specifically, the interplay of number patterns in rotation angles and their coefficients provides unified and intuitive generalized solution for the bidirectional pixel-to-pixel field transformation and reveals how light regulates complex amplitude in cascaded  bulk and metasurface WPs; addressing an elusive problem of polarization manipulation. In this study, we confined ourselves to validate the bidirectional complex amplitude modulation, parametric extraction, channels manipulations, and Jones matrix elements swapping through the generalized solution. It is important to emphasize that the underlying principle is general and universally applicable across different EM regimes, ensuring that our findings are not limited to a specific wavelength and can be translated to desired frequency domains with appropriate scaling. Additionally, sophisticated wavefront engineering applications such as holographic information encryption, vortex beam manipulation, and reconfigurable Moire's lensing etc., will emerge from this approach and can be achieved with known pixel-to-pixel complex amplitude information by translating and/or rotating the layers. We envision that the generalized approach of bidirectional field transformation and parametric extraction will serve as a guide to design new photonic devices and will expand the scope of polarization optics/photonics where the demand for compact, efficient, reconfigurable, and versatile optical components is rapidly growing.

\begin{figure} 
	\centering
	\includegraphics[width=0.95\textwidth, height=1.1\textwidth]{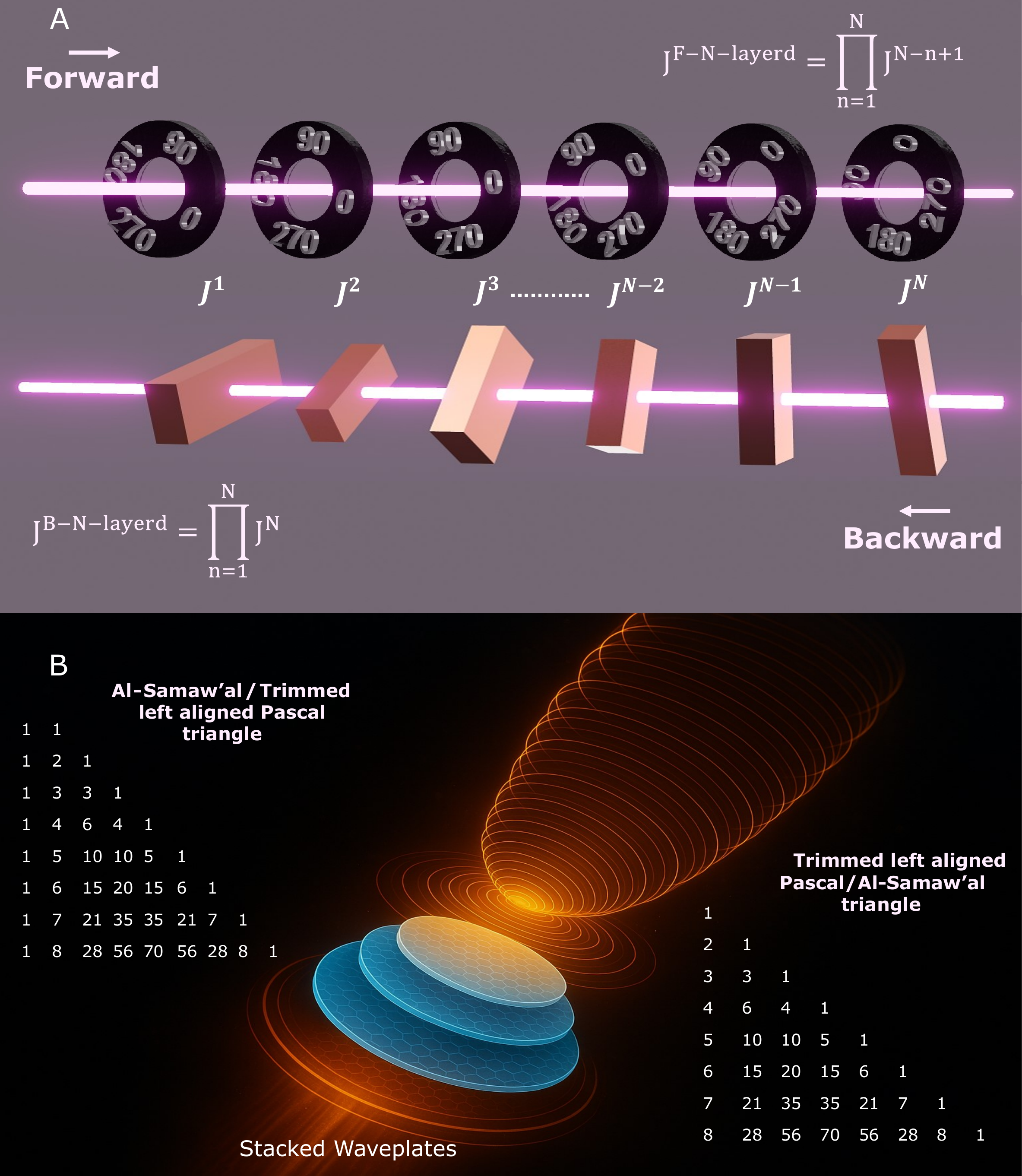} 

	\caption{\textbf{
		Generalized bidirectional Jones matrices solution for multilayer framework of twisted birefringent optical components and number pattern.} (\textbf{A}) Schematic principle of bidirectional complex amplitude modulation through in-plane rotated bulk and metasurface (rotated bricks) WPs arranged in cascaded configuration. (\textbf{B}) The numbers in the triangles are the binomial coefficients for stacked birefringent WPs ($N=1-8$).}
	\label{fig:Schematic}
\end{figure}

\clearpage 

%
\bibliography{Bib_Main} 
\bibliographystyle{sciencemag}

%
%
%
%
%
%


\section*{Acknowledgments}
All authors fully acknowledge that mathematical framework of this article is formulated by A.U.R.K. We acknowledgments the technical support of Kieran Rainey from EEECS, Queen's University, Belfast in mmWave experiments. And, thanks to Yessenia Jauregui-Sanchez for providing technical support in manuscript writing. The Dall-E was used for background image generation in Fig.\ref{fig:Schematic}b, which is just for pictorial demonstration only and doesn't reflect any result and finding. The Supplementary Video animations were created using the Manim animation engine.
\paragraph*{Funding:}
 The research was financially supported by the United Kingdom Research and Innovation (UKRI) Strength In Places Fund (SIPF) SmartNanoNI project. The Work of M.A.B.A was in part supported by the Department of Science Innovation and Technology (DSIT) project HASC Hub, and the Department for Digital, Culture, Media and Sport (DCMS) project REASON. The work of N.L and H.L was in part supported by National Natural Science Foundation of China (62405203); Key Discipline Construction Project of Guangdong Province (2022ZDJS110); Key Colleges and Universities Project of Guangdong Province (2023ZDZX1019).
\paragraph*{Author contributions:}
A.U.R.K developed the project and R.M.B supervised the  project. A.U.R.K conducted the numerical modeling and experiments and wrote the manuscript. N.U performed part of the simulation work and manuscript writing. N.U, N.L, H.L contributed to the data analysis and discussion. M.A.B.A supported the mmWave experiments. A.U.R.K, N.U., M.A.B.A, R.M.B revised the manuscript. All authors discussed the results and the final version of the manuscript.
\paragraph*{Competing interests:}
There are no competing interests to declare.
\paragraph*{Data and materials availability:}

The simulation results can be generated using the numerical methods described within Methods and Supplementary Information. The supplementary codes (Data \textcolor{blue}{S1} and \textcolor{blue}{S2}) used in this study will be publicly accessible through repository with doi at Zenodo platform after the peer-review. Additional data needed to evaluate the conclusions in the paper is presented in the paper and/or the supplementary materials.  \textbf{The supplementary information (Page \textcolor{blue}{S5}-\textcolor{blue}{66}) which supports the main text and other findings of this study can be obtained from the author.}




\subsection*{Supplementary materials}
Materials and Methods\\
Supplementary Text\\
Figs. \textcolor{blue}{S1} to \textcolor{blue}{S28}\\
Tables \textcolor{blue}{S1} and \textcolor{blue}{S2}\\
References \textit{(\textcolor{blue}{49}-\arabic{enumiv})}\\ 
Data \textcolor{blue}{S1} and \textcolor{blue}{S2}\\
Movies \textcolor{blue}{S1} to \textcolor{blue}{S3}\\


\newpage


\renewcommand{\thefigure}{S\arabic{figure}}
\renewcommand{\thetable}{S\arabic{table}}
\renewcommand{\theequation}{S\arabic{equation}}
\renewcommand{\thepage}{S\arabic{page}}
\setcounter{figure}{0}
\setcounter{table}{0}
\setcounter{equation}{0}
\setcounter{page}{1} 


\begin{center}
\section*{Supplementary Materials for\\ \scititle}

Ata~Ur~Rahman~Khalid$^{\ast}$, 
Naeem~Ullah, Nannan Li, Hui Li,\\ Muhammad Ali Babar Abbasi, Robert M. Bowman$^{\ast}$
 \\
\small$^\ast$Corresponding author. Email: ataurrahman.khalid$@$qub.ac.uk;r.m.bowman$@$qub.ac.uk\\
\end{center}

\subsubsection*{This PDF file includes:}
Materials and Methods\\
Supplementary Text\\
Figures S1 to S28\\
Tables S1 and S2\\
Captions for Data S1 to S2\\
Captions for Movies S1 to S3


\newpage


\subsection*{Materials and Methods}
\subsubsection*{Optical setup and polarization measurement}
The bulk optical WPs used in this study were procured from Thorlabs. The experimental setup of the cascaded bulk WPs for measuring the bidirectional polarization transformation is shown in Fig. S2. The setup was carefully aligned to ensure accurate measurements. A light beam with a wavelength of $630~nm$ was generated from a laser source (Thorlabs ITC4001-Benchtop Laser Diode/TEC Controller, 1A/96W) and initially directed by a fiber to the linear polarizer to generate linearly polarized light. The linearly polarized light was then passed through a different stacked WPs (Thorlabs), which were rotated at different angles to modulate the polarization state. The resulting beams were collected through polarimeter (Thorlab PAX1000IR1/M), and the obtained polarization states were recorded from Thorlab polarimeter software. The graphical data corresponding to different rotated stacked configurations were saved and plotted for further interpretation.
\subsubsection*{Fabrication and measurement in mmWave}
Two sets of samples consist of $15\times15$ anisotropic metalements rod array with a rotation $0:45:135^\circ$ (totaling 8 samples), and an isotropic square array tilted with an angle $67.5^\circ$ were fabricated using the standard printed circuit board (PCB) technique. The designs consist of $0.813~mm$-thick dielectric substrates with a dielectric constant of $3.38$ and copper-cladding of thickness $0.035~mm$ (RO4003C) which was etched to form the meta-element patterns. The total thickness of single fabricated samples is about $0.85 mm$. For the non-monolithic cascaded design, we punched holes in the coroners of PCB layers to allow layers stacking using nylon bolts integrated in home build sample holder as per design requirement. A $5~mm$ of fixed air gap was placed through the nylon nuts between each layer to achieve the desired EM response. The measurements of the fabricated samples were conducted in an anechoic chamber to minimize external interference. The horn antennas operating at a wideband of linearly polarized ``$7$ to $14~GHz$'' were employed as the feeding source to transmit and receive the EM waves to measure the sample's response in the microwave regime. The distance between the transmitting and receiving antennas was fixed at $d>20\lambda_o$; where $\lambda_o=13~GHz$. The transmitting and receiving antennas were connected to an Agilent Technologies 8381C vector network analyzer (VNA), which was utilized to measure the polarization response of the samples in forward and backward directions. In measurements, we took advantage of non-monolithic design approach by picking and placing the rotated layers on the top of the fixed layer. The photographs of fabricated samples and measurements are shown in Fig. S2.
\subsubsection*{Numerical simulations}
In optical regime, the numerical simulations were conducted using the commercially available \textsc{Ansys Lumerical FDTD Solution} software to model the behavior of the design in the optical domain. The numerical simulations were realized by silicon (Si) nanopillars/bricks to introduce complex amplitude transformation. The sub-wavelength lattice constant of the unit cell was set to $400~nm$ to ensure accurate modeling of the optical response. In unitcell simulation, periodic and perfectly matched layers (PML) boundary conditions were applied to minimize reflections. For OAM generation design, the optimized Si nanopillars/bricks were spatially placed on the SiO2 substrate. And, PML boundary conditions were applied to all directions
For unitcell simulations, the mesh size of $5~nm$ were utilized in ($X\&Y$)-directions and $20~nm$ in $z$-direction. The mesh size was modified to $25~nm$ in ($X,Y\&Z$)-directions for array designed for OAM. The operating wavelength was set at $700~nm$.

In mmWave regime, \textsc{CST Studio Suite} was utilized as a simulations tool for EM response analysis and parametric optimization. Periodic boundary conditions were employed in lateral directions and Floquet ports were used for excitation. The optimized unitcell periodicity ``P'' in lateral and vertical direction was $10~mm$., the optimized length and width of anisotropic copper rod was $8\times1.2~mm$, and the optimized side dimension of isotropic square patch was $6~mm\times6~mm$. The rogger (RO4003C) was used as a substrate material.
\subsubsection*{Data processing}
The analytical solution and raw data from numerical simulation and experiments were post-processed in \textsc{Matlab2022a}.
\subsubsection*{Details of experiments and measurement setup in visible and mmWave regimes}
\label{Sec:Exp_Setups}
The optical setup for the measurement of the polarization transformation using the generic cascaded WP scheme is shown in Fig. \textcolor{blue}{S1A} for $N=3$ layers. For other configurations, the optical WPs were placed and removed as per the design requirements. The optical measurement setup consists of a laser source, a linear polarizer (LP), the generic wave plates configurations under test, a polarimeter and a computer for analyzing the polarization states. The computer setup which is connected to the polarimeter is shown in Fig. \textcolor{blue}{S1B}. 

To measure the electromagnetic response of the mmWave metasurfaces, different prototypes of the fabricated metasurfaces and the mmWave measurement setup are shown in Fig. \textcolor{blue}{S2}. Different views and configurations of the fabricated isotropic and anisotropic metasurface samples are presented in Fig. \textcolor{blue}{S2A} with their description. For the non-monolithic cascaded design, we punched holes in the coroners of PCB layers to allow layers stacking using nylon bolts as shown in Fig. \textcolor{blue}{S2A} (the right most column) where the bi-layer and tri-layer non-monolithic cascaded designs are separated by nuts creating a fixed air gap of $5mm$. The
Fig. \textcolor{blue}{S2B} shows different steps of the measurement environment and the entire setup within a microwave anechoic chamber. For measurements, the metasurface samples are stacked as per design requirement in the home-build sample holder with the help of bolts and placed it centrally between the transmitting (Tx) and receiving (Rx) horn antennas. The electromagnetic response at the desired frequency was measured through the vector network analyzer.

\textbf{The supplementary information (Page S5-S66) which supports the main text and other findings of this study can be obtained from the author.\\
Supplementary Text\\
Figures S1 to S28\\
Tables S1 and S2\\
Captions for Data S1 to S2\\
Captions for Movies S1 to S3}


\end{document}